\documentclass[a4paper]{article}
\usepackage{cite, graphicx, subfigure, amsmath, appendix} 
\usepackage{setspace}
\usepackage{fullpage}
\usepackage{fancyhdr}
\usepackage{fullpage}
\usepackage{caption} 
\usepackage{hyperref}
\usepackage{multirow}
\usepackage{comment}
\usepackage{wrapfig}
\usepackage{amssymb}
\usepackage{amsfonts}
\usepackage{hyperref}
\usepackage{gensymb}
\usepackage{xcolor}
\usepackage[british]{babel}
\usepackage{hhline}
\usepackage{multirow}
\usepackage{adjustbox}
\date{}

\begin{document}
\newcommand\bra[2][]{#1\langle {#2} #1\rvert}
\newcommand\ket[2][]{#1\lvert {#2} #1\rangle}
\title{An approach to solve the coarse-grained Protein folding problem in a Quantum Computer}

\author{{Jaya Vasavi P}$^1$, {Soham Bopardikar}$^2$, {Avinash D}$^1$, {Ashwini K}$^1$, \\ {Kalyan Dasgupta}$^3$, {Sanjib Senapati}$^1$\\
$^1${Dept. of Biotechnology, Indian Institute of Technology Madras, Chennai, India}\\
$^2${Dept. of Electronics and Telecommunication Engineering, College of Engineering Pune, India }\\
$^3${IBM Research, Bangalore, India}}

\maketitle

\begin{abstract}
   Protein folding, which dictates the protein structure from its amino acid sequence, is half a century old problem of biology. The function of the protein correlates with its structure, emphasizing the need of understanding protein folding for studying the cellular and molecular mechanisms that occur within biological systems. Understanding protein structures and enzymes plays a critical role in target based drug designing, elucidating protein-related disease mechanisms, and innovating novel enzymes. While recent advancements in AI based protein structure prediction methods have solved the protein folding problem to an extent, their precision in determining the structure of the protein with low sequence similarity is limited. This limitation highlights the significance of investigating the thermodynamic landscape of proteins by sampling the diverse conformations they can adopt and identifying the global minimum energy structure. Classical methods face challenges in generating extensive conformational samplings, making quantum-based approaches advantageous for solving protein folding problems. In this work we developed a novel turn based encoding algorithm that can be run on a gate based quantum computer for predicting the structure of smaller protein sequences using the HP (Hydrophobic and Polar bead) model as an initial framework, which can be extrapolated in its application to larger and more intricate protein systems in future. The HP model best represents a major step in protein folding phenomena - the hydrophobic collapse which brings the hydrophobic amino acid to the interior of a protein. The folding problem is cast in a 3D cubic lattice with degrees of freedom along edges parallel to the orthogonal axes, as well as along diagonals parallel to the axial planes. The optimization problem gives us a formulation that can go up to sixth order, but, with suitable modifications and assumptions, can be brought down to a second order formulation, making it a QUBO problem. While, the original formulation with higher order terms can be run on gate based quantum hardwares, the QUBO formulation can give results on both classical softwares employing annealers and IBM CPLEX as well as quantum hardwares. Details of the VQE algorithm that was used for solving this problem is presented. We also present results representing the hydrophobic collapse obtained from IBM quantum simulators and real machines, and compare them with the ones obtained from PyQUBO (simulated annealing) and CPLEX.
\end{abstract}

\section{Introduction}
Proteins are important cellular machinery which performs many tasks like functioning as enzymes for chemical reactions within our cell, transporters for molecules through membrane, receptors for cell signaling, regulators for hormones and many more. All of these functions are made possible due to their three dimensional folded structure also called as native state. Proteins are made up of 20 different amino acids each having different physical and chemical properties like size, charge, hydrophobicity etc. Protein folding is a fundamental biological process, involving the intricate transformation of a linear chain of amino acids into a three-dimensional structure, crucial for the protein's optimal functionality. Despite having ~250 million protein sequence entries in the UniProt database\cite{the_uniprot_consortium_uniprot:_2023}, only ~181k structures are available in RCSB PDB\cite{wwpdb_consortium_protein_2019}. Most drug-designing strategies are based on protein structure, making it necessary to predict unknown disease-associated protein structures from their sequences. Under certain disease conditions or in the presence of a mutation, the proteins can form alternative misfolded structures that are non-functional or have a tendency to aggregate together. These misfolded proteins can disrupt normal cellular function, leading to cell death and disease. The importance of studying protein folding is emphasized by the association of misfolded proteins with diseases, such as Alzheimer's disease, Parkinson's disease, and Cystic fibrosis. So, predicting the protein folded structure and protein folding mechanism can help devise novel therapeutic strategies to prevent or rectify protein misfolding, presenting potential avenues for treating these diseases. Consequently, the study of protein folding holds immense importance in unraveling various biological processes and advancing the development of treatments for diverse diseases.

Protein folding is a complex process that involves many variables, including the sequence of amino acids, the environment surrounding the protein, and the interactions between different parts of the protein. One of the primary challenges of simulating protein folding using classical computing is the sheer number of possible configurations that a protein can take, described by Levinthal’s paradox \cite{Levinthal_Para}. Despite significant progress in computational methods that include many advanced MD simulation techniques and even machine learning/deep learning, classical computing has not yet been able to fully simulate protein folding accurately. Currently, protein folding simulations using classical computing are limited by the time and computational resources required to simulate even small proteins accurately.

To overcome these limitations, researchers have begun to explore the use of quantum computing for studying protein folding. Quantum computers can take advantage of quantum mechanical properties such as quantum superposition, interference and entanglement to perform many calculations simultaneously, which can greatly speed up simulations. For protein structure prediction problem in a 2D square lattice, it is possible to achieve a quadratic speedup over a classical computer, using a quantum algorithm. The space complexity can also be reduced to polynomial space when compared to the exponential space required in classical computer\cite{wong_fast_2022}. For example a protein of N amino acid length will require $2^N$ possible conformations to be sampled in a square lattice, which quickly reaches the maximum memory limit of a classical computer. Whereas, as quantum computer requires only a polynomial space of $\mathcal{O}(N^3)$ for the same task.

This paper is organized as follows. In section 2, we discuss some important literature with respect to solving the protein folding problem in a quantum computing machine. We highlight how this work differs from the work done previously. In section 3, we explain our turn based encoding formulation in detail. Section 4 gives the implementation steps required to encode a classical optimization problem for a quantum hardware. Section 5 discusses the results from classical algorithms as well as from IBM quantum simulators and hardwares. Finally, in section 6, we give the important conclusions.

\section{Literature Survey}
The first set of papers that discussed quantum algorithms for the protein folding problem is based on mapping the locations of protein molecules in a 2D/3D lattice in a coordinate space \cite{Guzik_protein}. They make use of the HP (H - hydrophobic, P - polar) lattice model used in classical algorithms. The coordinate space is represented by the quantum states of the qubits. For an $n$ qubit system we will have $2^n$ states. For example, in a 2D system, every molecule has to be given a state corresponding to the two coordinates (e.g. $x$ and $y$). For an $N$ molecule system, we will require $N$ states for the $x$ coordinate and $N$ states for the $y$ coordinate for every molecule. For uniqueness in the coordinate space, we will need $\log _2 N$ bits in any given axis for every molecule. So, overall, we will need $2N\log _2N$ qubits to represent all the coordinates. For a $D$ dimensional lattice, the number of qubits required will be $DN\log _2N$. The approach discussed in \cite{Guzik_protein} uses an adiabatic quantum evolution process to solve for the minimum energy state. The Hamiltonian is constructed by incentivizing proximity of H type of molecules and penalizing overlap between any two molecules and ensuring continuity of the chain (coordinates of adjacent molecules are unit distance apart). The qubit requirement is based on the worst case consideration that the entire chain of protein molecules may lie in any one of the axis without any movement in the other axis (horizontal to any axis). Since native structures of protein molecules come in a highly folded state, this kind of a methodology is not able to make maximum use of the entire state-space. 

The disadvantages in using a coordinate based approach led to models that used the so called turn based encoding approach \cite{Babbush_2014}. The turn based encoding approach maps the turns taken while moving from one molecule to the next to quantum states. This is analogous to a self avoiding walk (SAW) setup. In a 3D dimensional lattice, we could have 6 turns (above, below, right, left, front, back) from any given given point in the lattice. A 3-qubit encoding will suffice to represent all the movements. For an $N$ bead (molecule) system, we will have $N-1$ turns, and the total number of qubits required will be $3(N-1)$. A further assumption on taking the first turn as given will result in $3(N-2)$ qubit requirements. This is an improvement as far as qubit requirement is concerned. However, these methods are unable to capture diagonal movements and they also need ancilla qubits for encoding the slack variables required for enforcing the overlap constraints. In \cite{Babbush_2014}, the authors also discuss turn-circuit constructions that do not require ancilla qubits. These methods require very involved circuit constructions with higher order many-body terms. These kind of approaches have been tried out in quantum devices using annealing technologies. \cite{Babbush_2014} also discusses several ways to solve protein folding problems by mapping them to  quadratic unconstrained binary optimization (QUBO) problem, heuristic satisfiability problems, etc. QUBO problems are essentially 2-local problems (all terms of the optimization problem involve two or less number of variables) and have been widely used in many applications involving quantum computing machines. QUBO formulations are also solvable by classical algorithms like simulated annealing and tools like IBM CPLEX.

We also see other variants of turn based modelling of the protein folding problem. In \cite{Babej_2018}, the authors use a one hot encoding method, where every molecule is represented by $6$ qubits. Every qubit represents a direction in a 3D lattice. A qubit taking a value $\ket{1}$ indicates a turn in that direction. The encoding proposed here requires $6N - 7$ qubits. The method uses the QAOA approach to get to the ground state of the diagonal Hamiltonian \cite{Farhi_QAOA}. Some novelty may be needed in preparing the mixing Hamiltonian terms for the QAOA algorithm. The algorithm presented in \cite{Babej_2018}, however, takes into account movements only parallel to the axes and does not consider movements along the diagonals of the 3d cubic lattice. The QAOA approach would also require the exponentiation of the Hamiltonian resulting in large number of CNOT gates. This kind of an approach may not be easily implementable in present day quantum computers.
In \cite{Anton_2021}, the authors presented a turn based, resource efficient quantum model for the coarse-grained protein folding problem. The lattice considered in this work is tetrahedral in structure. The Hamiltonian had essentially three terms, mapping the geometrical constraints, the chirality constraints and the term corresponding to the interaction energies among amino acid molecules. This work considered the Miyazawa-Jernigan (MJ) interaction model for formulating the interaction Hamiltonian \cite{MIYAZAWA1996623}. The quantum algorithm used was cVAR-VQE (Conditional Value-at-Risk-Variational Quantum Eigensolver) \cite{Barkoutsos-cVar}. The authors also discuss sparse encoding methods to reduce quantum resource usage. 

The models discussed above were formulated keeping superconducting quantum machines in mind where the number of qubits available are limited. The usage of quantum annealers in the protein folding problem can also be found in the literature. The models formulated for quantum annealers require a large number of qubits. In \cite{Anders_2022}, the authors discuss a method using the lattice based HP model for quantum annealers. In this model, a qubit (variable) is assigned for every adjacency possible in a 2D lattice. A lattice with $L^2$ lattice sites and $N$ beads (molecules) would need $NL^2$ qubits. To reduce resource usage, the authors also propose an odd-even segregation of lattice points to reduce the qubit requirement by half. This scales as $\mathcal{O}(N^3)$ and are not feasible in superconducting quantum hardware. The Hamiltonian terms in this model map the overlap constraints (no two beads are at the same location), constraints that prevent the same molecule or bead from appearing at multiple locations in the lattice and finally the energy function that incentivizes adjacency of H beads. 

In this paper, we are proposing a turn based model in a 3D lattice for the protein folding problem. The peptide chain is coarse grained into a C-alpha backbone model where each bead represents the C-alpha atom of individual amino acids in the peptide sequence. Further, the interaction model could either be in the classical HP form, where only H beads have interaction, or it could be in the MJ form where interaction strengths of all beads are considered. With $N$ beads we will have $N-1$ turns, and every turn requires $6$ qubits for a 3D lattice. The number of qubits required in total is $6(N-1)$. We could follow the assumption given in \cite{Babbush_2014}, and consider the first turn as given. The number of qubits required would then be $6(N-2)$. Although the method proposed here uses more resources in terms of qubits, it provides for increased degrees of freedom in terms of the turns the amino acid chain can take. The method allows for diagonal movements both in a given plane (diagonal in a 2D square) as well as along a steric diagonal (the longest diagonal in a 3D cube). The chain while going from one bead to the next is capable of taking $6$ turns along the usual three axes ($\pm x, \pm y$ and $\pm z$), $12$ turns along the diagonals in the three axial planes ($4$ each in the $xy$, $yz$ and $zx$ planes), and $8$ turns along the steric diagonals. Overall, the degrees of freedom available is $26$. For a C-alpha backbone model on this 3D lattice, the internal angles can take 45\degree, 54.74\degree, 90\degree, 125.26\degree, 135\degree or 180\degree and these are possible between any two consecutive backbone bonds. The pseudo-angles between three consecutive C-alpha atoms in an actual protein varies between 80\degree to 155\degree\cite{boomsma_full_2005,oldfield_analysis_1994}, which is effectively covered by this 3D lattice. It is also important to note that the HP model of protein in cubic 3D lattice has a shortcoming termed as the parity problem - it is impossible to have contacts between two H beads if they are both in even or odd positions in the sequence. By extending the 3D lattice to include the diagonal this problem is effectively solved. The formulations result in the presence of up to 6-local terms (terms involving 6 variables), but we show that with suitable addition of constraints, we can limit the terms to 2-locals (QUBO formulation), without compromising on the degrees of freedom. The Quantum algorithm we have used is VQE. We also tried out the usage of cVARs as was done in \cite{Anton_2021}. We used Qiskit runtime estimators as well as samplers (sampling VQE) to get the desired bit-strings \cite{qiskit}. The quantum algorithms were run on both simulators as well as real IBM Quantum hardwares \cite{ibm-quantum}. The maximum number of beads we have used so far is $20$, requiring a total of $114$ qubits. We used 127 qubit IBM quantum hardwares to run our algorithms. We compare results given by quantum algorithms with the ones given by classical simulated annealing algorithms and IBM CPLEX.

\section{Turn based encoding formulation}
In this section we explain the proposed turn based encoding formulation and give the consolidated objective function. The original formulation has $n$-local ($n\geq 2$) terms. We also give the modified formulation with only 2-local terms (QUBO format). With a QUBO formulation, the optimization problem can be solved by classical as well as quantum algorithms.
\subsection{Variables and their symbols}
\label{var_symb}
\begin{itemize}
\item $N$ - Number of beads/amino acids
\item $S_H$ - Set of beads of the type $H$
\item $x_a^i, ~ x_b^i, ~ y_a^i, ~ y_b^i, ~ z_a^i, ~ z_b^i$ - Binary variables: $\{0,1\}$, $\forall ~ 1 \leq i \leq N-1$.
\begin{flalign}
\begin{aligned}
x_t^i = x_a^i - x_b^i \\ 
y_t^i = y_a^i - y_b^i \\ 
z_t^i = z_a^i - z_b^i 
\end{aligned}\label{turn_eqn}
\end{flalign}
$(x_t^i, ~ y_t^i, ~z _t^i)$ denote the turns taken in the three dimensions $(x, y, z)$ from the $i^{th}$ bead to the $(i+1)^{th}$ bead. With $N$ beads there can be only $N-1$ sets of $(x_t^i, ~ y_t^i, ~z _t^i)$. 
As a consequence of (\ref{turn_eqn}), we have the following expressions.
\begin{equation}
    \begin{array}{c}
        x_t^i, ~ y_t^i, ~z _t^i \in \{-1,0,1\} \\
        \left(x_t^i\right)^2, \left(y_t^i\right)^2, \left(z _t^i \right)^2 \in \{0,1\} 
    \end{array} \bigg\} ~~~ i = 1, 2, \hdots, N-1 \label{xt_range}
\end{equation}

\item $w_{jk}$ - Interaction weights between the $j^{th}$ and $k^{th}$ beads $\forall ~ j, k \in S_H$. 
Number of possible interactions - ${M \choose 2} = \frac{M!}{(M-2)!2!} = \frac{M(M-1)}{2}$, where $M = \#S_H$ (cardinality of $S_H$). We have to, however, exclude adjacent or bonded H beads from this list of interactions.
\end{itemize}

With $x_t^i, ~ y_t^i, ~z _t^i \in \{-1,0,1\}$, we can have the following turns, considering the origin, $(0,0,0)$, as reference.
\begin{enumerate}
    \item Turns parallel to the axes (6) - $(1,0,0), ~(-1,0,0), ~(0,1,0), ~(0,-1,0), ~(0,0,1), ~(0,0,-1)$ 
    \item Turns along a diagonal parallel to the axial planes
    \begin{itemize}
        \item $xy$ plane (4) - $(1,1,0), ~(-1,1,0), ~(-1,-1,0), ~(1,-1,0)$
        \item $yz$ plane (4) - $(0,1,1), ~(0,-1,1), ~(0,-1,-1), ~(0,1,-1)$
        \item $zx$ plane (4) - $(1,0,1), ~(-1,0,1), ~(-1,0,-1), ~(1,0,-1)$
    \end{itemize}

    \item Turns along the the steric diagonals - The set of all steric diagonals can always be divided into two parts: the first part above a given axial plane and the second part below the given plane. The plane could be any one of $xy$, $yz$ and $zx$. Given below are the parts as separated by the $xy$ plane. 
    \begin{itemize}
        \item above the $xy$ plane (4) - $(1,1,1), ~(-1,1,1), ~(-1,-1,1), ~(1,-1,1)$
        \item below the $xy$ plane (4) - $(1,1,-1), ~(-1,1,-1), ~(-1,-1,-1), ~(1,-1,-1)$
    \end{itemize}
\end{enumerate}
Overall, we can have a total of $26 ~(6+12+8)$ turn options or degrees of freedom. 

\subsection{Objective Function}
\begin{equation}
Obj = \sum_{j,k \in S_H} w_{jk}\left[ \left(\sum_{l=j}^{k-1} x_t^l\right)^2 + \left(\sum_{l=j}^{k-1} y_t^l\right)^2  + \left(\sum_{l=j}^{k-1} z_t^l\right)^2 \right] \label{obj-main}
\end{equation}
The objective function in (\ref{obj-main}) is the sum of all Euclidean distances among H beads. The weights $w_{jk}$ give priority based on the interaction strengths. For a purely $HP$ kind of formulation, the weights are all equal to $1$, i.e., $w_{jk} = 1, \forall ~j, k \in S_H$. 

\subsection{Continuity constraint}
Continuity constraint ensures that there is at least one turn in any one of the three dimensions between adjacent beads. In other words, not all of $(x_t^i, ~ y_t^i, ~z _t^i)$ equal 0.

\begin{equation}
    C_1 = \sum_{i=1}^{N-1} \left[ 1 -  \left(x_t^i\right)^2 -  \left(y_t^i\right)^2 -  \left(z_t^i\right)^2 +  \left(x_t^i\right)^2 \left(y_t^i\right)^2 + \left(y_t^i\right)^2 \left(z_t^i\right)^2\ + \left(z_t^i\right)^2 \left(x_t^i\right)^2 - \left(x_t^i\right)^2 \left(y_t^i\right)^2\left(z_t^i\right)^2 \right] \label{continuity_const}
\end{equation}
$C_1$ in (\ref{continuity_const}) will return a $1$ (high) when $(x_t^i, ~ y_t^i, ~z _t^i)$ equals $(0,0,0)$. It will return a $0$ (low), when any one or any two or all of $(x_t^i, ~ y_t^i, ~z _t^i)$ does not equal $0$. 

The condition when all of the turns equal $1$, i.e., $(x_t^i, ~ y_t^i, ~z _t^i) = (1,1,1)$, results in a 3-dimensional (3D) diagonal movement, also known as the steric diagonal movement. In literature we do find mentions of steric diagonals being avoided when a turn is considered \cite{Hans-2006}. To disincentivize that kind of a movement, we have to omit the last term $\left(x_t^i\right)^2\left(y_t^i\right)^2\left(z_t^i\right)^2$, as given in (\ref{continuity_const_steric}). This will give a high when $(x_t^i, ~ y_t^i, ~z _t^i)$ equals $(0,0,0)$ or $(1,1,1)$. 

\begin{equation}
    C_{1s} = \sum_{i=1}^{N-1} \left[ 1 -  \left(x_t^i\right)^2 -  \left(y_t^i\right)^2 -  \left(z_t^i\right)^2 +  \left(x_t^i\right)^2 \left(y_t^i\right)^2 + \left(y_t^i\right)^2 \left(z_t^i\right)^2\ + \left(z_t^i\right)^2 \left(x_t^i\right)^2 \right] \label{continuity_const_steric}
\end{equation}

\subsection{Overlap constraint}
The continuity constraint as explained above ensures that no two adjacent beads overlap. The more generic overlap constraint should ensure that no two non-adjacent beads have the same set of coordinates. This can be ensured by constraining the squared distance between all non-adjacent beads, in any one dimension, to be greater than $0$. This is as given in (\ref{overlap_const}).
\begin{equation}
    C_2 = \sum_{i=1}^{N-2} \sum_{j=i+2}^{N} \left[ \alpha _{ij}\left(\sum_{l=i}^{j-1} x_t^l\right)^2 + \beta _{ij}\left(\sum_{l=i}^{j-1} y_t^l\right)^2  + \gamma _{ij}\left(\sum_{l=i}^{j-1} z_t^l\right)^2 \right] \label{overlap_const}
\end{equation}
$\alpha _{ij}, \beta _{ij}$ and $\gamma _{ij}$ are chosen randomly such that any one of them is $1$ and the other two are $0$. One way of doing it is by randomly choosing three numbers from a normal distribution - $a, b, c$ and then assigning values to $\alpha _{ij}, \beta _{ij}$ and $\gamma _{ij}$ based on the following rules in (\ref{abc_random_rule}).
\begin{flalign}
\begin{aligned}
    a>b, ~c \implies \alpha _{ij} = 1, \beta _{ij} = \gamma _{ij} = 0 \\
    b>a, ~c \implies \beta _{ij} = 1, \alpha _{ij} = \gamma _{ij} = 0 \\
    c>a, ~b \implies \gamma _{ij} = 1, \beta _{ij} = \alpha _{ij} = 0 \\
\end{aligned} \label{abc_random_rule} 
\end{flalign}
This exercise has to be carried out for all pairs of $(i,j)$. This will penalize all non-adjacent beads from having equal sets of coordinates. This kind of a constraint (non-overlap in any one dimension) ensures minimum contradiction with the main objective function, which tries to bring the beads closer. There is an element of randomness here in selecting the dimension over which the non-overlap happens and this will also get reflected in the final result. To account for this inherent randomness in a classical setting, the best outcome could be selected from a population of sample runs. The equivalent Hamiltonian operator (of the best outcome) could then be forwarded to be run on a quantum hardware.

\subsection{Constraint to prevent diagonal crossing}
\begin{figure}[h]
\centering 
\includegraphics[width=5in]{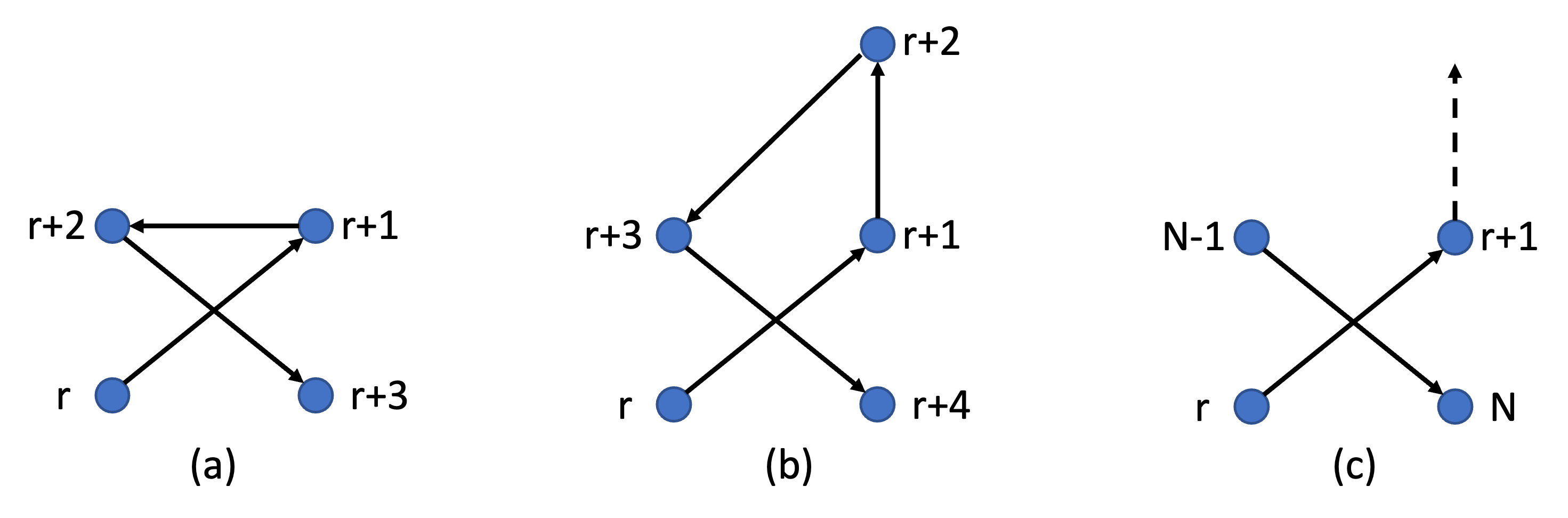} 
\caption{Diagonal crossing of bonded pairs}
\label{fig:diag_cross}
\end{figure}

In a valid conformation, crossing of two binding edges is forbidden \cite{Hans-2006}. Crossing of binding edges essentially imply diagonal crossing of bonded pairs. To prevent diagonal crossing of bonded pairs, we consider one bead at a time starting from $r=1$ (the first bead). We will look at all nearest possible diagonal crossing of bonded pairs that the $r^{th}$ bead can witness. The nearest possible diagonal crossings will happen when some bonded pair $(r+j,r+j+1)$ crosses the bonded pair of $(r,r+1)$. The first such instance can be that of, when, $j=2$. This is when the bond between $(r+2,r+3)$ crosses the bond between $(r,r+1)$ as depicted in Fig. \ref{fig:diag_cross}(a). The next instance could be when $j=3$, as shown in Fig. \ref{fig:diag_cross}(b). This could go on till we have the pair $(N-1,N)$ cross $(r,r+1)$, as shown in Fig. \ref{fig:diag_cross}(c).

Let $(x_r, y_r, z_r)$ be the coordinates of the $r^{th}$ bead. Also, let $k=r+j$. For the bonded pair $(r+j,r+j+1)$ (or $(k,k+1)$) to diagonally cross the pair $(r,r+1)$ at the centre (1:1 ratio), the following conditions must be met (section formula from coordinate geometry).
\begin{equation}
    \begin{array}{c}
        x_k + x_{k+1} = x_r + x_{r+1} \\
        y_k + y_{k+1} = y_r + y_{r+1} \\
        z_k + z_{k+1} = z_r + z_{r+1}
    \end{array}\Bigg\} ~~~
    \begin{array}{c}
    r = 1, 2, \hdots, N-3 \\ 
    k = r+2, \hdots N-1
    \end{array}\label{turn_diag_eqn}
\end{equation}
Now,
\begin{equation}
    \begin{aligned}
        x_k = \sum_{j=1}^{k-1} x_t^j, ~~~~ x_{k+1} = \sum_{j=1}^{k} x_t^j \\
        x_r = \sum_{j=1}^{r-1} x_t^j, ~~~~ x_{r+1} = \sum_{j=1}^{r} x_t^j.
    \end{aligned}\label{diag_coords}
\end{equation}

From (\ref{turn_diag_eqn}) and (\ref{diag_coords}) we have,
\begin{equation}
    \begin{aligned}
        (x_k + x_{k+1}) - (x_r + x_{r+1}) &= \left(\sum_{j=1}^{k} x_t^j + \sum_{j=1}^{k-1} x_t^j\right) - \left(\sum_{j=1}^{r} x_t^j + \sum_{j=1}^{r-1} x_t^j\right) \\
        &= \left(x_t^k + 2\sum_{j=1}^{k-1} x_t^j\right) - \left(2\sum_{j=1}^{r} x_t^j - x_t^r \right) \\
        &= x_t^k + x_t^r + 2\sum_{j=r+1}^{k-1} x_t^j \\
        &= X_{rk}
    \end{aligned} \label{rk_crossing}
\end{equation}

We will have a similar thing with $Y_{rk}$ and $Z_{rk}$. Only if all of $X_{rk}, ~Y_{rk}$ and $Z_{rk}$ equals zero, we will have a diagonal crossing between pairs $(k,k+1)$ and $(r,r+1)$. If any one of $X_{rk}^2, ~Y_{rk}^2, ~Z_{rk}^2 > 0$, a diagonal crossing is avoided. We use a similar concept as in $C_2$.

\begin{equation}
    C_3 = \sum_{r=1}^{N-3} \sum_{k=r+2}^{N-1} \left[ \alpha _{rk}X_{rk}^2 + \beta _{rk}Y_{rk}^2 + \gamma _{rk} Z_{rk}^2\right]
\end{equation}
$\alpha _{rk}, \beta _{rk}$ and $\gamma _{rk}$ are chosen randomly such that any one of them is $1$ and the other two are $0$. The methodology to choose could be similar to the one used in $C_2$.

\subsection{Modifications to reformulate as a QUBO problem}
Except for the higher order terms in (\ref{continuity_const_steric}), all other terms in the rest of the objective function and constraints have 2-local terms or less. Let us have a closer look at (\ref{continuity_const_steric}).
\begin{equation*}
    C_{1s} = \sum_{i=1}^{N-1} \left[ 1 -  \left(x_t^i\right)^2 -  \left(y_t^i\right)^2 -  \left(z_t^i\right)^2 +  \left(x_t^i\right)^2 \left(y_t^i\right)^2 + \left(y_t^i\right)^2 \left(z_t^i\right)^2\ + \left(z_t^i\right)^2 \left(x_t^i\right)^2  \right] 
\end{equation*}
From (\ref{turn_eqn}), the terms of the type $\left(x_t^i\right)^2$ can essentially be written as follows.
\begin{equation}
\begin{aligned}
    \left(x_t^i\right)^2 &= (x_a^i - x_b^i)^2 =  \left(x_a^i\right)^2 + \left(x_b^i\right)^2 - 2x_a^ix_b^i \\
    &= x_a^i + x_b^i - 2x_a^ix_b^i 
\end{aligned} \label{xt_2_exp}
\end{equation}
In (\ref{xt_2_exp}) above, since the variables $x_a^i, x_b^i \in \{0,1\}$, we have used the relations $\left(x_a^i\right)^2 = x_a^i$ and $\left(x_b^i\right)^2 = x_b^i$.
All the terms in (\ref{xt_2_exp}) are 2-locals or less. Now, let us look at the terms of the type $\left(x_t^i\right)^2 \left(y_t^i\right)^2$. We will omit the superscript $i$ for ease of writing. 
\begin{equation}
\begin{aligned}
    \left(x_t\right)^2 \left(y_t\right)^2 &= (x_a - x_b)^2(y_a - y_b)^2 \\
    &= x_ay_a + x_ay_b + x_by_a + x_by_b - 2x_ax_b(y_a + y_b) - 2y_ay_b(x_a + x_b) + 4x_ax_by_ay_b
\end{aligned} \label{xt_yt_2_exp}
\end{equation}
The first four terms in (\ref{xt_yt_2_exp}) are of the type $x_ky_l, ~k,~l \in \{a,b\}$. These are two local terms. The three local terms are of the type $x_ax_b(y_a + y_b)$ and the 4 local term is of the type $x_ax_by_ay_b$. In either case, the terms will exist only when we have situations where $x_a = x_b = 1$ (for the 3-local term) and $x_a = x_b = 1,~ y_a = y_b = 1$ (for the 4-local term).

Let us now look at all the values that the variable $x_t$ takes for different values of $x_a$ and $x_b$. Table \ref{table:xt_table} gives the values. The table will remain the same for $y_t$ and $z_t$.
\begin{table}[!h]
\caption{Values of $x_t$ for different values of $x_a, x_b$}
\label{table:xt_table}
\begin{center}
\begin{tabular}{|c|c|c|}
 \hline
 $x_a$ & $x_b$ & $x_t = x_a - x_b$ \\
\hline
0 & 0 & 0 \\
0 & 1 & -1 \\
1 & 0 & 1 \\
1 & 1 & 0 \\
\hline
\end{tabular} 
\end{center}
\end{table}

$x_t$ can take the values $\{-1, 0,1\}$ as given in Table \ref{table:xt_table} and as mentioned in (\ref{xt_range}). Clearly, the cases where $x_a = x_b = 0$ and $x_a = x_b = 1$ return the same value of $x_t = 0$. As such, even if we avoid the situation where $x_a = x_b = 1$, the range of $x_t$ is not diminished. However, on avoiding the state, $x_a = x_b = 1$, the 3-local and 4-local terms in (\ref{xt_yt_2_exp}) can be avoided. On avoiding the terms $x_a = x_b = 1$, $y_a = y_b = 1$ and $z_a = z_b = 1$, the constraint equation in (\ref{continuity_const_steric}) gets modified to the following equation.
\begin{equation}
    C_{1s}^{mod} = \sum_{i=1}^{N-1} \left[ 1 -  \left(x_t^i\right)^2 -  \left(y_t^i\right)^2 -  \left(z_t^i\right)^2 +  \sum_{k,l=a,b}x_k^iy_l^i + \sum_{k,l=a,b}y_k^iz_l^i + \sum_{k,l=a,b}z_k^ix_l^i\right] \label{continuity_const_steric_mod}
\end{equation}
where $C_{1s}^{mod}$ is the modified form of $C_{1s}$.

Now, to avoid situations like $x_a = x_b = 1$, $y_a = y_b = 1$ and $z_a = z_b = 1$, we have to add another constraint as given in (\ref{xa_xb_const}).
\begin{equation}
    C_4 = \sum_{i=1}^{N-1} x_a^ix_b^i + y_a^iy_b^i + z_a^iz_b^i \label{xa_xb_const}
\end{equation}

\subsection{Consolidated Objective function with penalties}
The consolidated objective function with all the penalty functions can be expressed as follows.
\begin{equation}
    \min_{x_a^i, x_b^i,y_a^i,y_b^i,z_a^i,z_b^i} \lambda _0.Obj + \lambda _1.C_{1s}^{mod} - \lambda _2.C_2 - \lambda _3.C_3 + \lambda _4.C_4 ~~~~ i = 0,1, \hdots, N-1 \label{eqn:final_form}
\end{equation}
$\lambda _0, ~\lambda _1, ~\lambda _2, ~\lambda _3, ~\lambda _4 \geq 0$, are the appropriate weighting/penalty coefficients for the objective and constraint functions. The penalty factors are chosen based on the impact it has on the consolidated objective function.

\subsection{Total Energy Calculation}
\begin{figure}[htp]
\centering 
\includegraphics[width=6in]{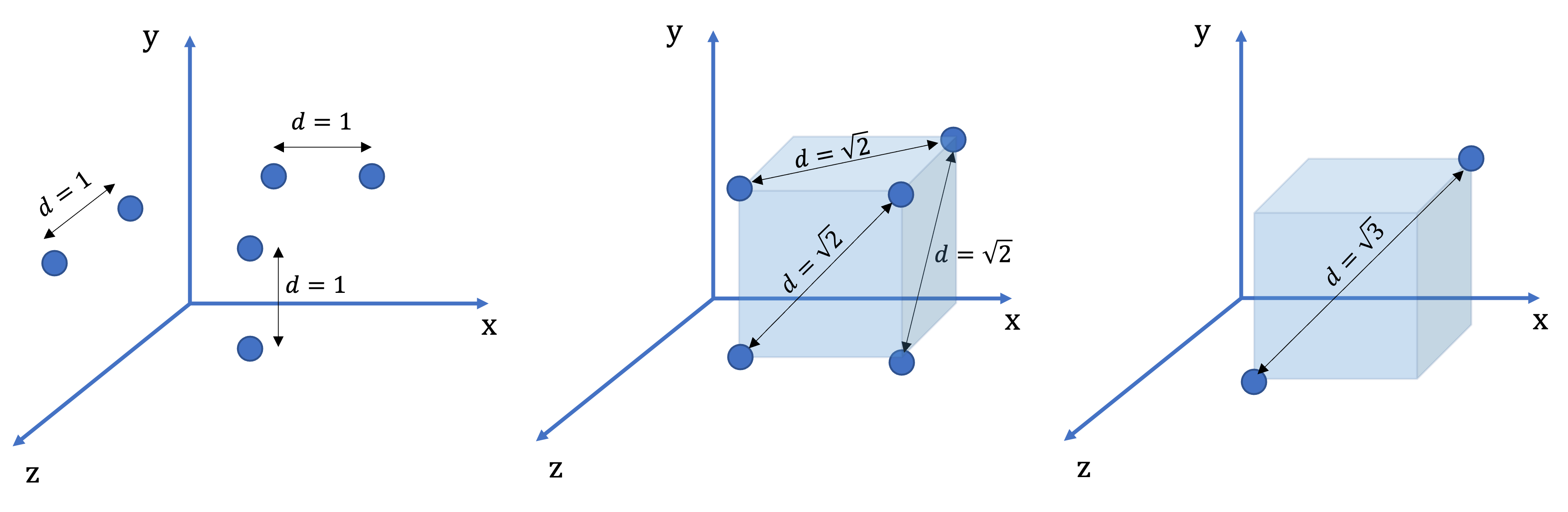} 
\caption{Distances between H beads that are in proximity}
\label{fig:proximity}
\end{figure}

In an HP configuration, the total energy is calculated based on the number of non-adjacent H beads that are in proximity to each other. Adjacent H beads in the chain are anyway in proximity, and they are not counted. We consider it a proximity when two H beads are within a unit lattice distance away in all 3 spatial dimensions. The shortest distance would be a one unit distance away, and that would happen when they share the same coordinates in any two dimensions. We could also have distance of $\sqrt{2}$, when they share the same coordinate in any one dimension and are a unit distance away in the other two dimensions. The maximum distance that two beads, that are in proximity, can have is $\sqrt{3}$. This happens when they are a unit distance away in all three dimensions. Fig. \ref{fig:proximity} illustrates the idea.

Every time we get a proximity between two non-adjacent H beads we add a $-1$ to the total energy. The best result in a search space will be the one that gives the least value of energy with all constraints satisfied. It is to be noted that the energy value we are discussing here is different from the minimized consolidated objective function value. The hope is that the minimization of the consolidated objective function value will lead to a maximization of the adjacency of H beads.

\subsection{Selection of weighting/penalty factors}
If we look at the components in the formulation in (\ref{eqn:final_form}), we will see that the components pertaining to the objective function, the overlap constraints and the constraints to prevent diagonal crossings have a measure of euclidean distances. The overlap constraints ($C_2$) and the constraints to prevent diagonal crossings ($C_3$) have a negative sign in front of them. The weighing or the penalty factors are $\lambda _2$ and $\lambda _3$ respectively. The weighing factor for the main objective function, $Obj$, is $\lambda _0$. Any optimization algorithm would work to strike a balance between $Obj$ on one side and $C_2$ and $C_3$ on the other side. If the weighing factors make the constraint side heavier, we will have no violations (all constraints satisfied) at the cost of a relatively higher energy value. On the other hand, if the weighing factors make the main objection function heavier, we will get lower energy values but with violations. Striking a balance should ideally give us both, i.e., lower energy values with no violations. 

We know from section \ref{var_symb} and from the expression in (\ref{obj-main}), that the number of terms in $Obj$, is $n_0 = {M \choose 2} - Adj = \frac{M(M-1)}{2} - Adj$, where $Adj$ indicates the number of bonded or adjacent H beads. From the expression in (\ref{overlap_const}), the number of terms involving overlap constraints is 
\begin{equation}
    n_2 = (N-2) + (N-3) + \hdots + 2 + 1 = \frac{(N-2)(N-1)}{2}
\end{equation}
Similarly, from (\ref{turn_diag_eqn}), the number of terms involving diagonal crossing constraints is
\begin{equation}
    n_3 = (N-3) + (N-4) + \hdots + 2 + 1 = \frac{(N-3)(N-2)}{2}
\end{equation}
An ideal case for the weighing factors should be such that the following expression is true.
\begin{equation}
    \lambda _0 n_0 \approx \lambda _2 n_2 + \lambda _3 n_3 \label{eqn:weigh_expr}
\end{equation}
In majority of the cases in a HP kind of interaction model, we will have $n_2 > n_0$, unless $M = N-1$. Similarly, we will have $n_3 \geq n_0$, unless $M>N-2$. Given such circumstances, we are most likely to have $ \lambda _0 > \lambda _2$. We can fix $\lambda _2 = 1$ and $\lambda _3$ to a relatively small value (some trials required here) and calculate $\lambda _0$ from (\ref{eqn:weigh_expr}). We can try out higher values of $\lambda _0$ to push up the number of adjacent H beads (in an HP model) or lower the overall energy value (in an MJ model), but we need to ensure that no violations are present. This requires fine adjustment of $\lambda _0$ around the number calculated in (\ref{eqn:weigh_expr}).

The weighing factors $\lambda _1$ and $\lambda_4$ are for logical (output of $0/1$) set of constraints  $C_{1s}$ and $C_4$. These factors can be kept at a large value compared to the other penalty factors and may require some experimentation on setting the value.

\section{Implementation on a Quantum computer}
\begin{figure}[htp]
\centering 
\includegraphics[width=6in]{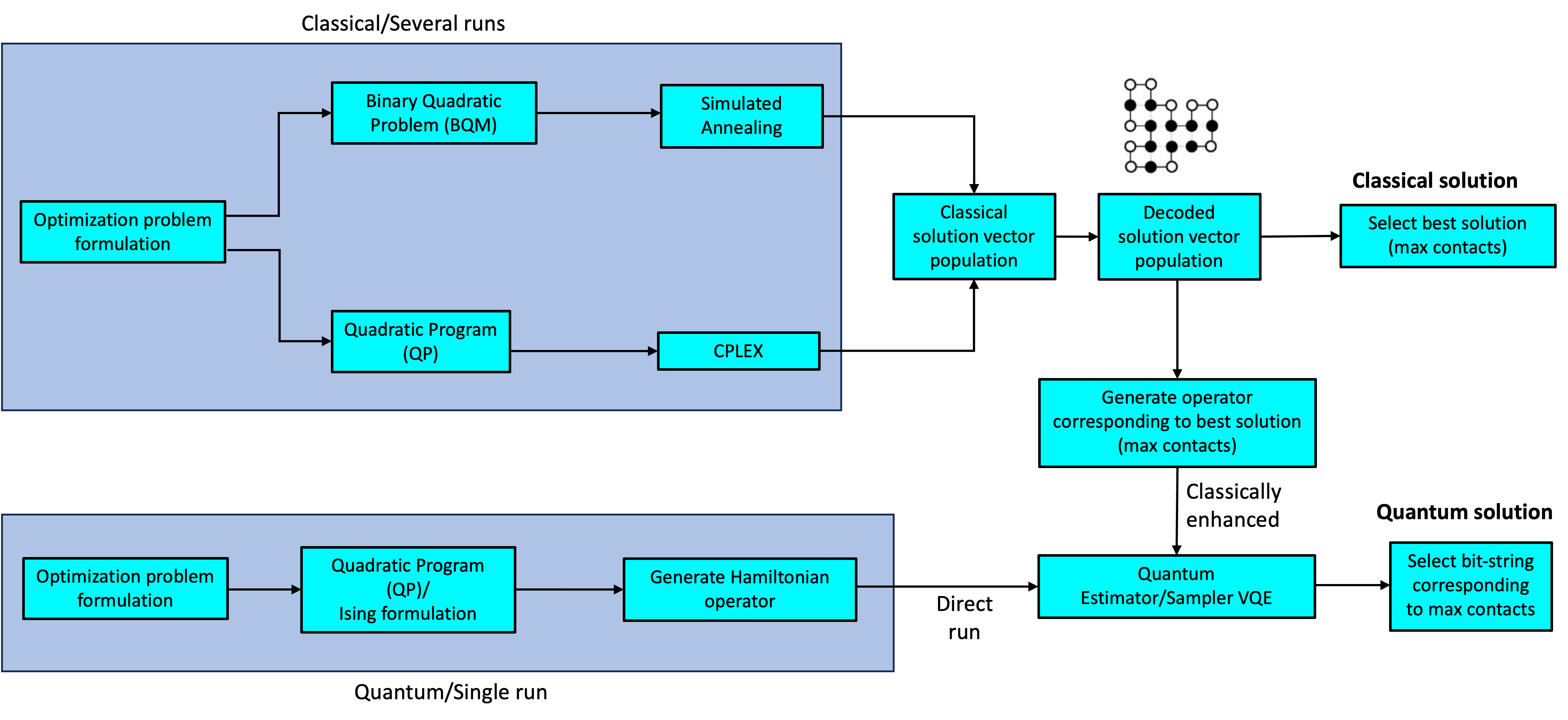} 
\caption{Steps for solving an optimization problem on a quantum computer}
\label{fig:flowchart}
\end{figure}

There are several steps to solving the optimization problem as shown in Fig. \ref{fig:flowchart}. In this section, we will discuss the process of solving the optimization problem on IBM's superconducting (gate based) quantum computers, and bench-marking them with quantum simulators and classical tools like CPLEX and Simulated Annealing. 

To run it in a Quantum hardware or simulator using quantum algorithms, we need to generate the Hamiltonian operator from the formulation. The detailed steps are discussed in the following sections. As discussed above, there is randomness in the formulation when the overlap and diagonal crossing constraints are taken into account. To get the best results, we can solve the optimization problem multiple times and select the best outcome from a population of solution vectors. Doing the same exercise in a Quantum hardware is not feasible, given the limited hardware time available. To get the best results in a Quantum hardware, we could use the Hamiltonian operator corresponding to the best classical outcome. We would call this classically enhanced, as shown in Fig. \ref{fig:flowchart}. One can also do a direct single run to generate the Hamiltonian operator from a one time formulation. 

\subsection{Solving the problem using classical algorithms}
The process starts with coding the turn-based encoding formulation in the form of an objective function along with constraints which is done with the help of the PyQUBO library. PyQUBO is an open-source Python library that offers the convenience of effortlessly generating QUBO models through flexible mathematical expressions. Key features of PyQUBO include its Python-based nature with a C++ back-end, automatic constraint validation, and provision for parameter tuning \cite{9369010}. This library provides a user-friendly approach to creating QUBOs, allowing researchers to focus on their research problem without worrying much about the code. 

QUBO as discussed before is a mathematical formulation used in quantum computing to solve optimization problems. It involves minimizing a quadratic polynomial over binary variables \cite{glover_qubo}. It is expressed as follows:-

\begin{equation}
\begin{aligned}
    \min x^{T}Qx + c^{T}x \\
    Q \in \mathcal{R}^{nxn}, c\in \mathcal{R}^{n} \\
\end{aligned} \label{eqn:qubo_form}
\end{equation}
where, $x$ is a vector of binary decision variables and $Q$ is a square matrix containing the coefficients. All constraints (quadratic or less) are considered as penalties and added to the objective function with suitable penalty factors. This augmented objective function takes the form of (\ref{eqn:qubo_form}). Section 3 has already provided a detailed description of the way we have formulated the turn-based encoding formulation in the form of a QUBO problem. 

Simulated annealing is a heuristic algorithm that explores solution spaces and finds the optimal solution, demonstrating its efficiency in solving a wide range of optimization problems. PyQUBO creates a \textit{model} object once the formulation is compiled. To run it on a simulated annealer we convert this to a Binary Quadratic Model (BQM) object, and run it on PyQUBO's in-built annealing samplers. 

We have also used IBM CPLEX to obtain classical results. IBM CPLEX is a powerful optimization tool widely used to solve a variety of complex problems. It is particularly effective in solving integer programming (IP) and mixed-integer programming (MIP) problems \cite{cplex2009v12}. Qiskit has libraries (\textit{QuadraticProgram}) in its optimization module that can convert \textit{model} objects to a form that can be read by CPLEX. 

The results from these two have been used as a benchmark to see the performance of the real quantum device for our problem.

\subsection{Solving using Quantum algorithms}
To map the QUBO formulation into a Quantum computer, we convert it into the Ising formulation. The goal is to find the Hamiltonian operator $F_{Ising}$ that encodes the cost function $F_{QUBO}$ \cite{10.3389/fphy.2014.00005}.

\begin{equation}
\begin{aligned}
    F_{QUBO} = \sum_{i,j=1, i \neq j}^{N}x_{i}Q_{ij}x_{j} + \sum_{i=1}^{N}c_{i}x_{i} \\
    \textrm{where ~~~} x_i, x_j \in \{0,1\} 
\end{aligned} \label{eqn:qubo}
\end{equation}
The Ising and QUBO models are related through the transformation $x_{i} = \frac{I-Z_i}{2}$.
\begin{multline}
    F_{Ising} = \sum_{i,j=1,i\neq j}^{N}\frac{1}{4}Q_{ij}Z_{i}Z_{j} - \frac{1}{4}\sum_{i=1}^{N}\left(2c_{i}+\sum_{j=1}^{N} Q_{ij}+\sum_{j=1}^{N} Q_{ji} + 2Q_{ii}
    \right)Z_{i} \\
    + \left(\sum_{i,j=1, i\neq j}^{N}\frac{Q_{ij}}{4}+\sum_{i=1}^{N}\left(\frac{c_{i}}{2} + \frac{Q_{ii}}{2}\right)\right)I ~~~~~~~ \textrm{where ~~} \lambda (Z_i),\lambda (Z_j) \in \{-1,1\}.
\label{eqn:ising}
\end{multline}
In \ref{eqn:ising}, $F_{QUBO}$ is the QUBO cost function, $F_{Ising}$ is the Ising Hamiltonian operator, $I$ is the Pauli $I$ matrix and $Z$ is the Pauli $Z$ matrix. The products of Paulis shown here are essentially tensor products. The last term on the right would be the coefficient of $I^{\otimes N}$.

Since, Qiskit does not directly support PyQUBO, we created a code converter that transforms the QUBO formulation in PyQUBO to an instance of a Quadratic Program in Qiskit. Using our PyQUBO plugin, we first convert the QUBO equation to the Ising Hamiltonian format. Finally, we run it on a real IBM quantum hardwares and simulators with VQE and CVaR algorithms to get the solution vector for our protein problem. 

The Variational Quantum Eigensolver (VQE) is a well known quantum algorithm that harnesses the power of quantum computing to approximate the ground state of quantum systems. Its application spans various fields, including quantum chemistry, materials science, and optimization. To enhance the efficiency and robustness of VQE, researchers have recently introduced Conditional Value-at-Risk (CVaR) optimization, a risk assessment measure commonly used in financial risk management.

\subsubsection{VQE Overview} \label{VQE_explain}
\begin{figure}[htp]
\centering 
\includegraphics[width=5in]{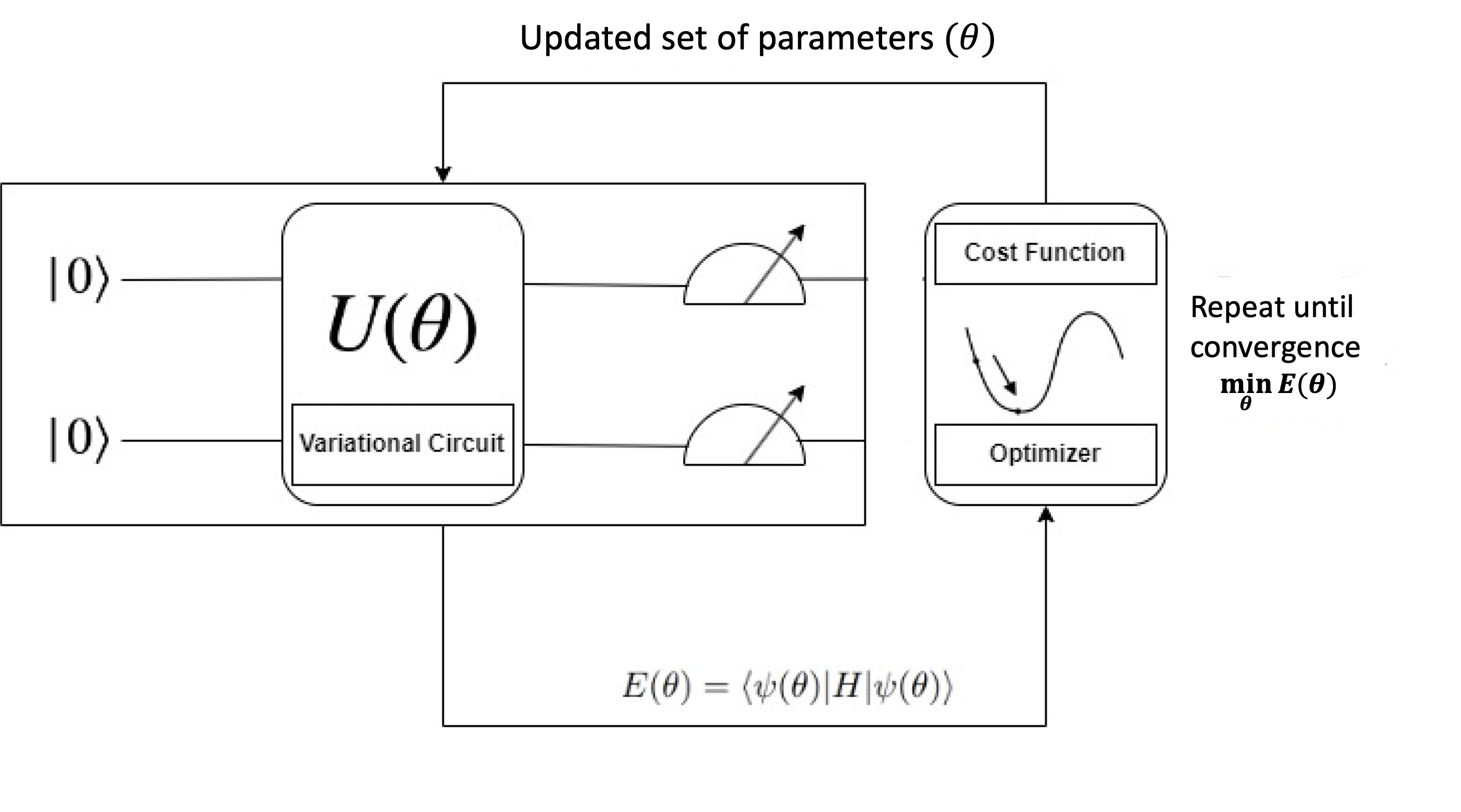} 
\caption{Schematic View of VQE Algorithm}
\label{fig:vqe}
\end{figure}

The VQE algorithm as shown in Fig. \ref{fig:vqe} starts by preparing an initial trial wavefunction, often referred to as the ansatz, which is a parameterized quantum state. The ansatz is typically chosen based on prior knowledge of the system or physical intuition. The parameters of the ansatz are then optimized classically to minimize the measured energy, following the variational principle. This optimization process is performed iteratively, adjusting the parameters of the ansatz to approach the optimal solution. During each iteration, the quantum circuit corresponding to the ansatz is executed on a quantum computer, and the expectation value of the Hamiltonian operator is calculated. The expectation value is essentially the energy of the trial wavefunction. The classical optimizer updates the parameters of the ansatz based on the measured energy, aiming to find the set of parameters that minimizes the energy\cite{grimsley2019adaptive}.

The VQE algorithm is designed to be compatible with near-term quantum devices, which are characterized by limited qubit coherence times and high error rates \cite{Moll.2018}. It is expected to have a significant impact on the development of quantum technologies and the exploration of quantum advantage over classical supercomputers \cite{McClean.2016}.

In summary, the VQE algorithm leverages the quantum-classical hybrid approach, wherein a quantum computer performs quantum operations, while classical optimization routines adjust the quantum parameters. It aims to minimize the energy expectation value of a given Hamiltonian, H, with respect to a parameterized quantum state, $\ket{\psi(\theta)}$. The parameter optimization process involves evaluating the energy expectation value, $E(\theta) = \langle \psi(\theta) \rvert H \rvert \psi(\theta) \rangle$ and updating the parameters to minimize the energy, thereby approximating the ground state \cite{Peruzzo}.

\subsubsection{Combining VQE with CVaR for optimization} \label{CVaR_explain}
Conditional Value-at-Risk (CVaR) is a risk measure that quantifies the potential risk beyond a given threshold. In the context of VQE, CVaR optimization allows us to account for the worst-case scenarios in estimating the ground state energy. By considering the tail of the energy distribution, CVaR provides a more comprehensive risk assessment than traditional mean-based measures \cite{Barkoutsos-cVar}. This is especially useful in cases where the Hamiltonian operator is diagonal, like we have in our case. Given a cumulative density function $F_X$ of a random variable $X$, the CVaR of $X$ is defined as the expected value of $X$ when $X<F^{-1}(\alpha)$ when the confidence interval is $\alpha$, i.e.


\begin{equation}
    CVaR_{\alpha}(X)=\mathbb{E}[X|X \leq F^{-1}_X(\alpha)], \alpha = (0,1]
\end{equation}
In other words, CVaR is the expected value of the lower $\alpha$-tail of the distribution of $X$. 

In a diagonal Hamiltonian, all bitstrings are eigenstates. It is also easy to calculate the energy value corresponding to these eigenstates since it involves only calculating the parity (eigenvalues are $+1$ and $-1$). Every circuit execution yields a bitstring/eigenstate which in turn is associated with an energy value. A normal VQE expectation is estimated by averaging the energy values over all circuit executions. In CVaR, on the other hand, the averaging happens the lower $\alpha$-tail. Without loss of generality, assume that the energy samples $E_{k}$ are sorted in non-decreasing order, the CVaR can be calculated as follows.
\begin{equation}
    CVaR_{\alpha}(\{E_1, E_2, ... , E_k\})=\frac{1}{[\alpha K]}\sum_{k=1}^{[\alpha K]}E_k.
\end{equation}
$\alpha = 1$ corresponds to the expected value one would normally get in a VQE operation. CVaR in general, has been used for a variety of different combinatorial optimization problems \cite{amaro2022case}, \cite{chai2023towards}. In our problem, we have used $\alpha = 0.05$ and $0.1$ while implementing CVaR. 

\subsection{Solution Vector Decoding}
\begin{figure}[htp]
\centering 
\includegraphics[width=6in]{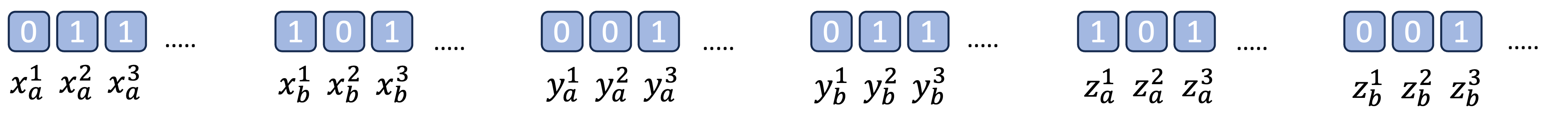} 
\caption{Example solution bitstring}
\label{fig:bitstring_expl}
\end{figure}

For an $N$ bead system, the final result is a bitstring of length $6(N-1)$, that gives the minimum expectation energy value. The bitstring will give the values of $x_a^i, ~ x_b^i, ~ y_a^i, ~ y_b^i, ~ z_a^i, ~ z_b^i$. The index $i$ will run from $1$ to $N-1$. The turns can then be calculated by using equation (\ref{turn_eqn}). Fig. \ref{fig:bitstring_expl} gives an example solution bitstring.

From the figure, one can clearly calculate that $x_{t}^1 = x_a^1 - x_b^1 = -1, ~~y_{t}^1 = y_a^1 - y_b^1 = 0$ and $z_{t}^1 = z_a^1 - z_b^1 = 1$. In other words, the first turn will be from $(0,0,0)$ to $(-1,0,1)$. One can, similarly, calculate the values of turns in the three axes for all the remaining indices.

\section{Results}
We employed Variational Quantum Eigensolver (VQE) to access the best quantum states. Our experiments were conducted on the 127 qubit IBMQ hardware (\textit{ibm}\_\textit{cusco}). We opt VQE over QAOA, a well known approach of Variational Quantum Algorithm for solving optimization problems, primarily due to QAOA’s increasing complexity with the rise in number of operator (Hamiltonian) terms, resulting in longer run times, as we increase the sequence length. The number of operator terms can easily run into thousands. Since, QAOA involves the evolution of the problem Hamiltonian, our circuit depth would become very large for any decent sized sequence length. This is not an ideal condition for present day Quantum hardwares where Quantum Volume (QV) is limited \cite{QV_Gambetta}. VQE, on the other hand does not have this issue as explained in section \ref{VQE_explain}.

\begin{figure}[h]
\centering 
\includegraphics[width=6in]{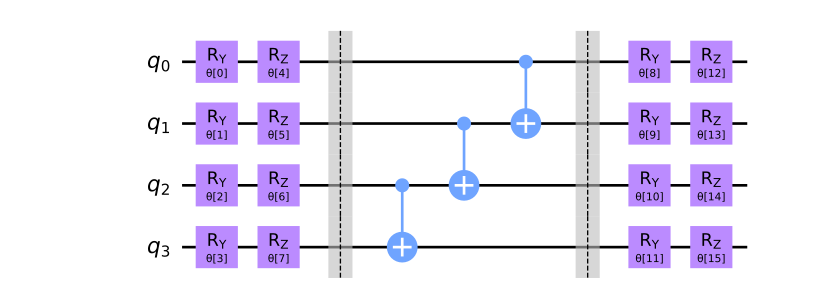} 
\captionof{figure}{A representation of the Efficient SU(2) ansatz for a 4 qubit system with one rotational - entanglement layer}
\label{su2}
\end{figure}

\begin{figure}[h]
\centering 
\includegraphics[width=6in]{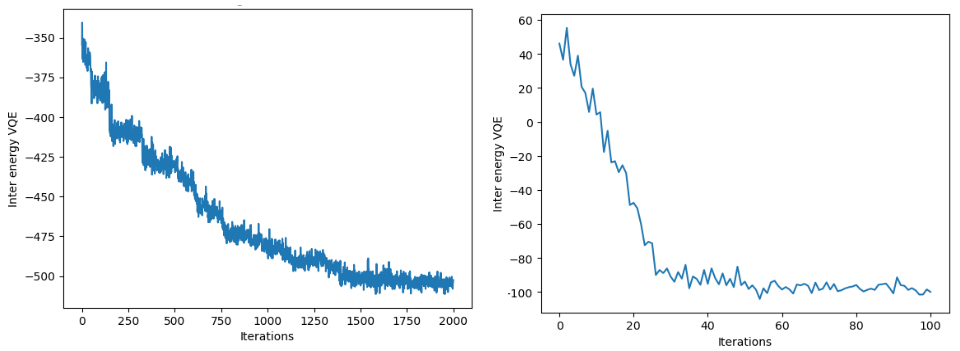} 
\captionof{figure}{Convergence of energy eigenvalue of a 10 bead sequence (HPPHPPHPHH) using VQE Sampler (left), VQE Estimator (right)}
\label{conv_sampl_vs_est}
\end{figure}

We utilized both VQE Sampler and VQE Estimator techniques, with an Efficient SU(2) ansatz (Fig. \ref{su2}) having one to two repetitions, thereby restricting depth and to the two-local terms. For the Estimator, using the optimized parameters that minimizes the expectation value of the Hamiltonian, we sampled the ansatz to access the quantum states (bitstrings). In the case of the Sampler, rather than optimizing the expectation value, we optimize the Conditional Value-at-Risk as introduced by \cite{Barkoutsos-cVar}. To see the convergence properties of the two different techniques, we initially ran them in Qiskit qasm simulator \cite{qiskit}. It was found that Sampler required iterations in the thousands for the 10 bead sequence to converge. The Estimator on the other hand, converged in a few hundred iterations, as shown in Fig. \ref{conv_sampl_vs_est}. We ran the Estimator in IBM hardware for as many number of iterations as it took to reach a saturation value of VQE energy in the simulator. For example, in Fig. \ref{conv_sampl_vs_est}, the Estimator primitive took around 100 iterations to come to a stop in qasm simulator, even though the saturation value was reached at around 50 iterations. The Estimator would then be run for 50 iterations in the IBM hardware. To expedite the process while using the Sampler primitive, we initialized the hardware with the parameters obtained from the simulator after convergence, and then ran the Sampler on the hardware for 50 more iterations. To mitigate the gate induced errors in VQE-Estimator and sampling error in VQE-Sampler, we used Zero-noise extrapolation (ZNE)  \cite{PhysRevLett.119.180509} and M3 \cite{mthree} techniques respectively by choosing the appropriate resilience level in Qiskit Runtime.

Considering the vast sample space – for instance the 10 bead sequence demands 54 qubits, resulting in $2^{54}$ possible quantum states, we selected the best 4000 quantum states by setting the number of shots to 4000. The bitstring that minimizes the expectation of the Hamiltonian operator should ideally give us the best conformation. However, we cannot be absolutely sure that our penalty factors are the best optimized. To be absolutely sure about the best conformation, we search through all the 4000 best states so obtained, and select the state that has no violation (all constraints satisfied) and has maximum number of contacts (or minimum energy value) as our final conformation. This ability to access so many states is what gives quantum hardware its real power. 

We decode obtained bitstrings to coordinate space, assessing our model by calculating non-adjacent hydrophobic bead contacts. The maximum possible contacts are computed as ${M \choose 2} - c = \frac{M(M-1)}{2} - c$, where $M$ is the hydrophobic bead count, and $c$ is the number of adjacencies among hydrophobic beads. The number of contacts were compared with the ones given by IBM Quantum simulators, as well as with the classical results from simulated annealers and IBM CPLEX as presented in the Table \ref{res}. We can see that the VQE consistently performs better. Notably, the VQE Estimator excels with fewer iterations, often outperforming VQE Sampler. It also appears that when there are more number of adjacent H beads, Estimator or Sampler performs better than the classical methods (last two sequences in Table \ref{res}). When there are multiple H beads that are adjacent, there are chances of the constraints, especially the overlap constraint, getting violated. Sampler and Estimator are able to give out options that can give a higher number of contacts without violating the constraints.

\begin{scriptsize}
\begin{table}[h]
  \hspace{-1.15cm}
  \centering
  \renewcommand{\arraystretch}{1.5}
  \begin{tabular}{|p{1.2cm}|p{3.0cm}|p{1.5cm}|p{1.3cm}|p{1.8cm}|p{1.5cm}|p{1.7cm}|}
    \hline
    \multirow{2}{1.2cm}{\textbf{Length}} & \multirow{2}{5cm}{\textbf{HP Sequence}}  & \multirow{2}{2cm}{\textbf{Max. contacts}} & \multicolumn{4}{c|}{\textbf{Results}} \\
    \cline{4-7}
    & & & \textbf{CPLEX} & \textbf{Annealing}& \textbf{Sampler} & \textbf{Estimator}\\
    \hline
    10 & PPHPPHPPHP & 3 &3 & 3& 3 & 3\\ \hline
    10 & HPPHPPHPHH & 9 & 9 & 9 & 9 & 9 \\ \hline
    10 & HHPPHPHPHP & 9 & 7 & 8 & 9 & 9  \\ \hline
    10 & HHHHPPHPHH & 17 &14 & 14 & 17 & 17  \\ \hline
    10 & HHHHHPHHHH & 28 &18 & 18 & 25 & 26  \\ \hline
  \end{tabular}
  \caption{Comparison of maximum number of contacts among classical (CPLEX, Simulated annealing) and Quantum hardware (VQE sampler, estimator) results for different 10 bead protein sequences}
  \label{res}
\end{table}
\end{scriptsize}
\begin{figure}[h]
\centering
\begin{minipage}{\textwidth}
  \centering
  \adjustbox{trim=0.2cm 0cm 0cm 0cm}{
  \includegraphics[width=1.1\textwidth]{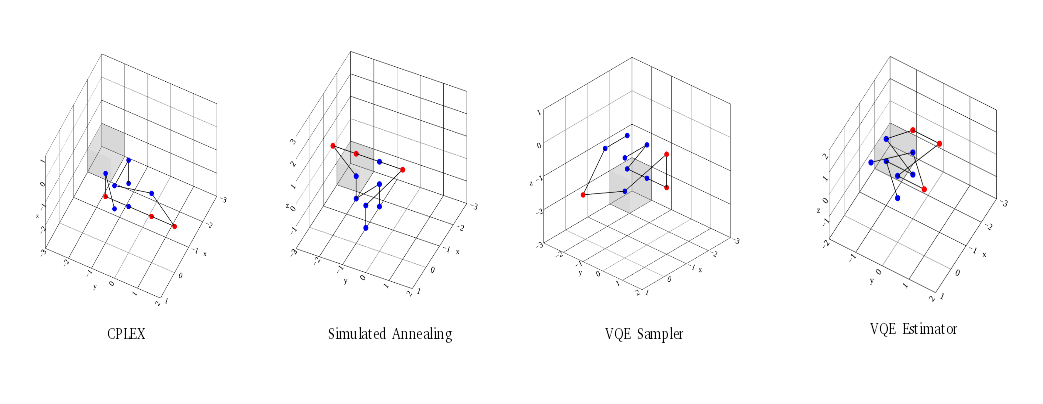} }
\end{minipage}

\captionof{figure}{Hydrophobic collapse of HHHHPPHPHH sequence among different optimizers}
\label{convPlots}
\end{figure}

Fig. \ref{convPlots} shows the hydrophobic collapse for a 10 bead sequence (HHHHPPHPHH) where the hydrophobic beads represented as blue, tends to come to the interior of the protein to form a hydrophobic core whereas the polar beads are on the periphery. Hydrophobic collapse is an essential phenomena that happens during the protein folding process. 

It was also found that the iterations that the VQE Estimator takes to reach the saturation value of VQE energy is dependent on the hydrophobicity (proportion of H beads) of the sequence. Table \ref{hydro_iter} gives the approximate number of iterations it took to reach saturation for different levels of hydrophobicity in $10$ bead systems. Clearly, there is a gradual increase in the number of iterations as the hydrophobicity increases. As the hydrophobicity increases, it becomes harder to reach the optimal solution. This is primarily due to the fact that the algorithm has to work harder to bring the large number of H beads closer without violating any of the constraints. Quantum algorithms, like VQE and other hybrid approaches, can excel in these kind of situations, as is evident from the results in Table \ref{res}. 
\begin{table}[h!]
\hspace{-0.8cm}
    \centering
    \begin{tabular}{|c|c|c|}
        \hline
       HP Sequence  &  Hydrophobicity (\%) & Iterations \\
       \hline
       PPHPPHPPHP & 30 & 50 \\
       HPPHPPHPHH & 50 & 50 \\
       HHPPHPHPHP & 50 & 150 \\
       HHHHPPHPHH & 70 & 170 \\
       HHHHHPHHHH & 90 & 250 \\
       \hline
    \end{tabular}
    \caption{Iterations to reach saturation value of VQE energy as a function of hydrophobicity}
    \label{hydro_iter}
\end{table}

\begin{figure}[h]
\centering 
\includegraphics[width=6in]{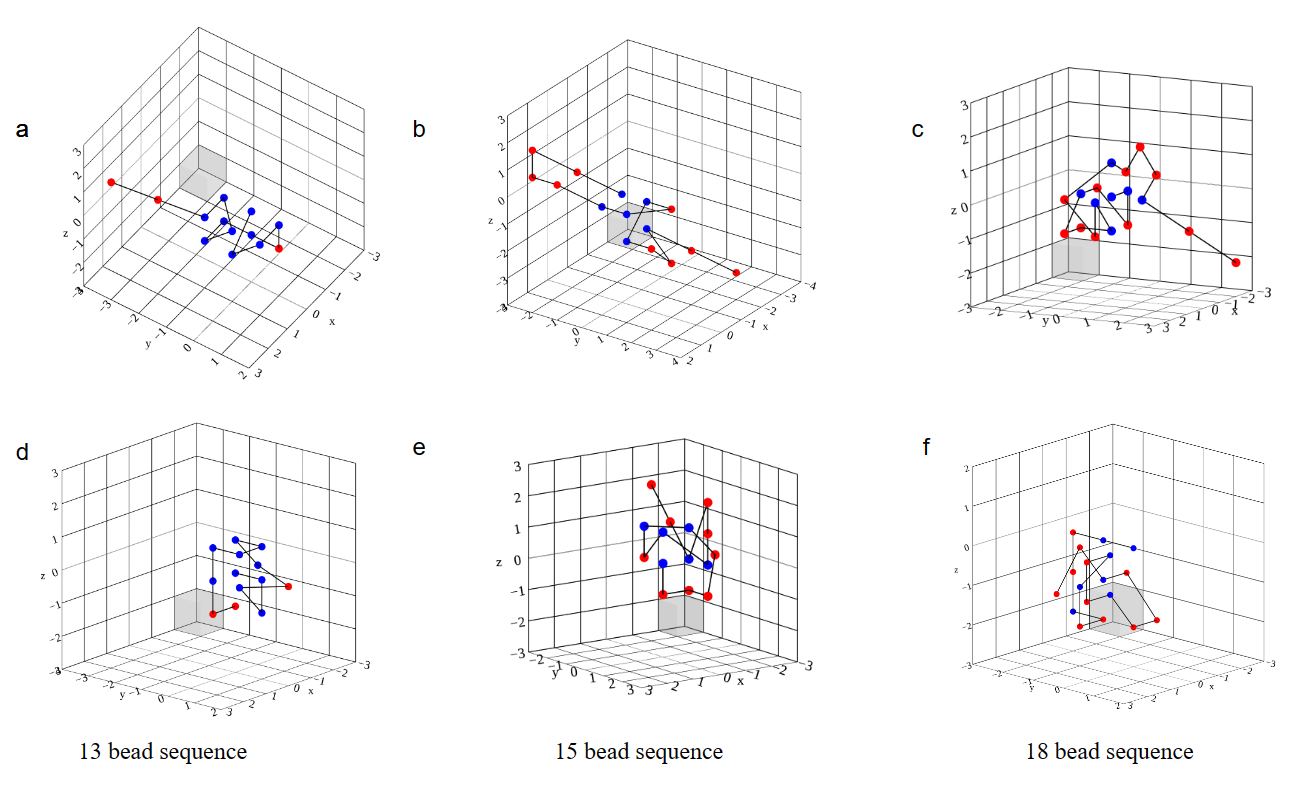} 
\captionof{figure}{a, b, c represents the Simulated annealing; and d,e,f represents the VQE estimator results respectively}
\label{convPlots1}
\end{figure}

Since the results generated by the VQE Estimator were close to the theoretical maximum of possible HH contacts, we proceeded to use the VQE estimator for longer protein sequences as presented in Table \ref{res1}. From Fig. \ref{convPlots1}, we can observe that both the classical (simulated annealing) and IBM hardware VQE Estimator are able to achieve hydrophobic collapse for longer protein sequences by bringing the hydrophobic beads together. The maximum number of HH contacts obtained through VQE Estimator are comparable to the classical results (Table \ref{res1}), and interestingly, structures obtained from IBM hardware are more compact than the ones obtained from classical simulated annealing.

\begin{table}[h!]
  \hspace{-0.8cm}
  \centering
  \renewcommand{\arraystretch}{1.3}
  \begin{tabular}{|p{1.2cm}|p{5cm}|p{1.5cm}|p{1.8cm}|p{1.8cm}|}
    \hline
    \multirow{2}{1.2cm}{\textbf{Length}} & \multirow{2}{3.1cm}{\textbf{HP Sequence}}  & \multirow{2}{2cm}{\textbf{Max. contacts}} & \multicolumn{2}{c|}{\textbf{Results}} \\
    \cline{4-5}
    & & & \textbf{Annealing} & \textbf{Estimator}\\
    \hline
    13 & HPPHPPHPHH & 37 & 29 &29 \\ \hline
    15 & HPPPPHHPHHPPHPP & 13 & 13 & 12 \\ \hline
    18 &HHPPHPPHHPPHPPPHPP & 19 & 16 & 15  \\ \hline
    20 &PHPHPHHPPHHHPPHPHHHP & 50 & 36 & 32  \\ \hline
   
  \end{tabular}

  \caption{Comparison of maximum number of contacts for longer protein sequences}
  \label{res1}
\end{table}

\break

\begin{figure}[h]
\centering 
\includegraphics[width=5in]{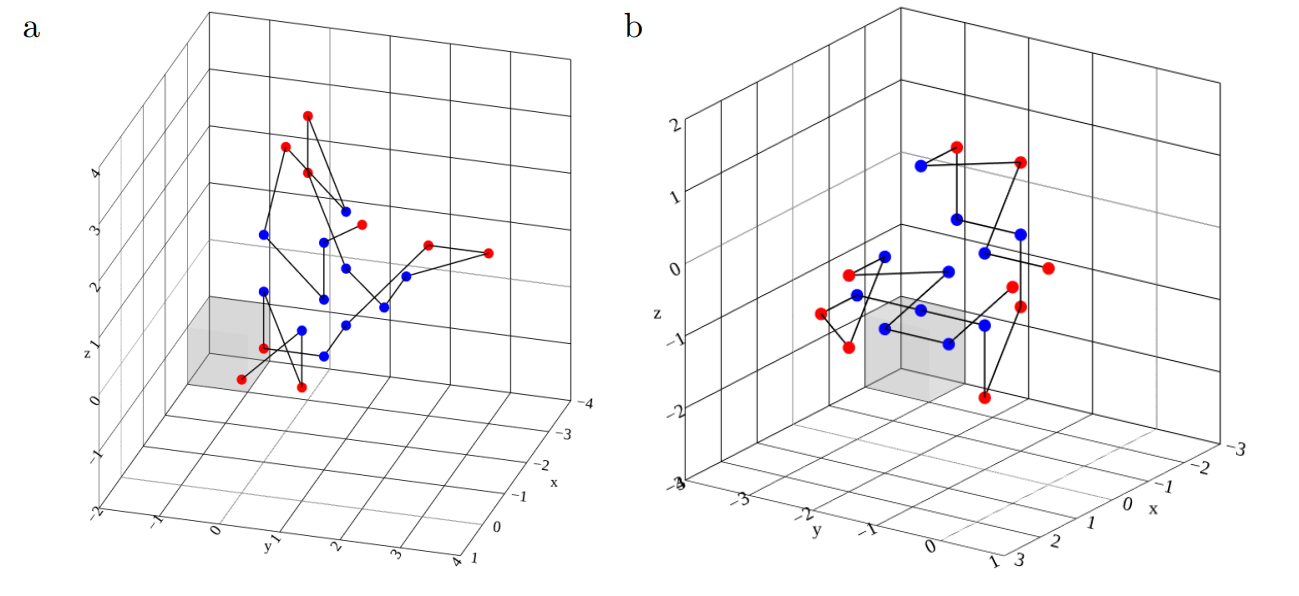} 
\captionof{figure}{a, b represents the simulated annealing and IBM hardware estimator results respectively for a 20 bead protein sequence PHPHPHHPPHHHPPHPHHHP}
\label{convPlots2}
\end{figure}

Fig. \ref{convPlots2}, depicts the folding of 20 bead amino acid sequence, which utilizes 114 qubits. 20 bead amino acid sequences are the longest sequences we have experimented with in a 127 qubit system. It can be clearly seen that the folding has happened without any overlaps or any diagonal crossings between bonded pairs. While our model accurately predicts the hydrophobic collapse for sequences having lesser number of contacts, it falls slightly short (when compared with the simulated annealer) as the sequence length and the number of contacts increases, failing to capture the collapse fully. 

\break

\section{Conclusions}
In this paper we have discussed a novel turn based protein folding formulation that can be run on Quantum computing hardwares using variational algorithms. The highlight of the formulation is in the increased degrees of freedom that a turn has while going from one protein molecule to the next. We ran our formulation using both classical optimization algorithms as well as quantum algorithms in machines from IBM Quantum, and the results were compared. Hydrophobic collapse, the most predominant factor in the phenomena of protein folding, could be captured in both Quantum and classical simulations. Quantum algorithms were found to give better results when the sequences had lesser number of contacts or when there were many adjacent H beads. For sequences having higher number of contacts, the classical algorithms gave slightly better results. We used both Sampler as well as Estimator primitives in Qiskit runtime to estimate the energy values. In conjunction with VQE, we also used the CVaR technique to zero down on the most ideal bitstring. We went as far as to use $114$ qubits in a $127$ qubit machine (\textit{ibm}\_\textit{cusco}) for a $20$ bead system. With a reasonably shallow circuit, it is possible to get a large number of meaningful outcomes in combinatorial optimization problems.

The ansatzes we used were very shallow in depth to prevent noise pollution. A future course of action would be to experiment with a variety of ansatzes with larger circuit depths, and see the effect it has on the folded structures. By successfully solving a higher-order optimization problem on a gate-based quantum computer, this paper not only challenges the conventional preference for quantum annealers but also opens up new avenues for leveraging the power of gate-based quantum computing in previously unexplored problem domains. Though this paper primarily demonstrates the hydrophobic collapse within an HP interaction model, the formulation can be easily extended to include other forms of interactions as well. This will also be the scope of our future work.

\bibliographystyle{IEEEtran}
\bibliography{reference}
\end{document}